\documentclass[sigconf,authorversion,nonacm]{acmart}
\setcopyright{acmcopyright}
\copyrightyear{2021}
\acmYear{2021}
\acmDOI{00.0000/0000000.0000000}

\acmConference[Preprint]{Preprint}{August 31, 2021}{}
\acmBooktitle{Preprint}
\acmPrice{15.00}
\acmISBN{XXX-X-XXXX-XXXX-X/XX/XX}

\usepackage{graphicx}
\usepackage{subcaption}

\usepackage[utf8]{inputenc}

\usepackage{balance}

\usepackage{amsfonts}
\usepackage{nicefrac}

\usepackage{amsmath}
\usepackage{xcolor}
\usepackage{paralist}
\usepackage{enumerate}

\usepackage{float}
\usepackage{bbm}
\usepackage{mathrsfs}
\usepackage[misc,geometry]{ifsym}

\usepackage{array}
\usepackage{booktabs}
\usepackage{multirow}
\usepackage{diagbox}

\usepackage{hyperref}
\usepackage{algorithm2e}

\usepackage{caption}

\begin{document}
\title{Dynamically Modelling Heterogeneous Higher-Order Interactions for
	Malicious Behavior Detection in Event Logs}
%
\author{Corentin Larroche}
\email{corentin.larroche@ssi.gouv.fr}
\affiliation{%
  \institution{French National Cybersecurity Agency (ANSSI)}
  \streetaddress{51, boulevard de La Tour-Maubourg}
  \city{Paris}
  \country{France}
  \postcode{75007}
}
\affiliation{%
  \institution{Télécom Paris}
  \streetaddress{19, place Marguerite Perey}
  \city{Palaiseau}
  \country{France}
  \postcode{91120}
}

\author{Johan Mazel}
\email{johan.mazel@ssi.gouv.fr}
\affiliation{%
  \institution{French National Cybersecurity Agency (ANSSI)}
  \streetaddress{51, boulevard de La Tour-Maubourg}
  \city{Paris}
  \country{France}
  \postcode{75007}
}

\author{Stephan Clémençon}
\email{stephan.clemencon@telecom-paris.fr}
\affiliation{%
  \institution{Télécom Paris}
  \streetaddress{19, place Marguerite Perey}
  \city{Palaiseau}
  \country{France}
  \postcode{91120}
}

\renewcommand{\shortauthors}{Larroche et al.}

\begin{abstract}
Anomaly detection in event logs is a promising approach for intrusion
detection in enterprise networks.
By building a statistical model of usual activity, it aims to detect multiple
kinds of malicious behavior, including stealthy tactics, techniques and
procedures (TTPs) designed to evade signature-based detection systems.
However, finding suitable anomaly detection methods for event logs remains an
important challenge.
This results from the very complex, multi-faceted nature of the data: event
logs are not only combinatorial, but also temporal and heterogeneous data,
thus they fit poorly in most theoretical frameworks for anomaly detection.
Most previous research focuses on one of these three aspects, building
a simplified representation of the data that can be fed to standard anomaly
detection algorithms.
In contrast, we propose to simultaneously address all three of these
characteristics through a specifically tailored statistical model.
We introduce \textsc{Decades}, a \underline{d}ynamic, h\underline{e}terogeneous
and \underline{c}ombinatorial model for \underline{a}nomaly
\underline{d}etection in \underline{e}vent \underline{s}treams, and we
demonstrate its effectiveness at detecting malicious behavior through
experiments on a real dataset containing labelled red team activity.
In particular, we empirically highlight the importance of handling the multiple
characteristics of the data by comparing our model with
state-of-the-art baselines relying on various data representations.

\end{abstract}


\maketitle

\section{Introduction}

Beyond rousing and slightly delusive forecasts of undefeatable
AI-powered intrusion detection systems, anomalous activity detection in
computer networks can be seen as the intersection of two main concepts.
The first one is behavioral analysis, which could be defined as
focusing on activity rather than static attributes – for instance, looking for
processes repeatedly encrypting files in a short time span rather
than executables matching known ransomware samples.
The second one is anomaly detection, an approach aiming to characterize the
benign rather than the malicious, defining the latter as "not known good"
instead of "known bad".
A practical example is looking for unusual parent-child
process associations rather than traces of a specific family of exploits.
These two paradigms share a common goal: building more generic and robust
detection methods,
capable of spotting advanced attackers using stealthy tactics, techniques and
procedures (TTPs) such as
living-off-the-land or supply chain attacks.
End goals of such threat actors include espionage, sabotage and financial gain.

In practice, behavioral models of computer networks are often built using
event logs, which are basically records of various kinds of activity happening
within a given perimeter.
These logs can be generated either by hosts or network equipments.
Being able to build a statistical model of events recorded inside a given
network theoretically allows
security experts to detect all kinds of malicious behaviors, as long as they
are inconsistent with what usually happens in the computer network at hand.
However, fully addressing the complexity of such data remains a major
challenge.
Indeed, let alone the sheer variety of benign and malicious behaviors that
must be dealt with, event logs do not easily lend themselves to
mathematical characterization.
They are primarily combinatorial data, in that each event is defined by a
finite number of fields mostly containing discrete values: user and host names,
IP addresses, and so on.
But there is also a temporal dimension: events appear at specific instants,
and they result from activity that evolves over time.
Finally, the multiplicity of actions occurring in a computer network makes
event logs intrinsically heterogeneous: for instance, it hardly makes sense
to compare an authentication and a process creation.

Due to this complexity, standard anomaly detection algorithms cannot be
used as such on event logs:
the significant semantic gap between the data and the model
must be bridged.
Most existing work circumvents this problem by adapting the data to the model,
simplifying them enough to make them fit into a standard anomaly detection
framework~\cite{yen2013beehive,legg2015automated,hu2017anomalous}.
This leads to significant information loss, in turn bounding the performance
of the whole detection procedure: once relevant aspects of the data have been
erased, even the smartest algorithm will be unable to
recover them.
Therefore, recent contributions have developed a more comprehensive
approach, coming up with better suited models instead of oversimplifying
the data~\cite{tuor2018recurrent,amin2019cadence,leichtnam2020sec2graph}.
In particular, significant progress has been made towards adequately modelling
the combinatorial dimension of events.
Our work builds upon these recent advances, extending them to simultaneously
address the temporal and heterogeneous aspects in addition to the combinatorial
one.

We propose \textsc{Decades}, a \underline{d}ynamic, h\underline{e}terogeneous
and \underline{c}ombinatorial model for \underline{a}nomaly
\underline{d}etection in \underline{e}vent \underline{s}treams.
This model can be used to learn the usual activity patterns in a computer
network, subsequently identifying anomalies at the event granularity.
We make use of multi-task learning to efficiently handle the heterogeneity of
event logs, improving our model by jointly learning
the behavior of entities across all event types.
We also design a retraining strategy to track the evolution of the network
and avoid concept drift, enabling fine-tuning of the
balance between fitting to new data and remembering past information.
Finally, we evaluate \textsc{Decades} against several state-of-the-art
baselines and show its effectiveness on a real dataset containing labelled red
team activity~\cite{kent2015cyberdata}.

The rest of this paper is structured as follows.
We first give some basic definitions and review existing work in
Section~\ref{sec:problem_formulation}.
The next two sections are devoted to describing our model:
we address the combinatorial and heterogeneous dimensions in
Section~\ref{sec:model}, and Section~\ref{sec:evolution} deals with the
updating procedure we use to handle temporal evolution of the data
distribution.
We present our experiments and results in Section~\ref{sec:experiments}.
Finally, we discuss some limitations of \textsc{Decades} along with leads for
future research in Section~\ref{sec:discussion} and briefly conclude in
Section~\ref{sec:conclusion}.

\section{Problem Formulation and Related Work}
\label{sec:problem_formulation}

We start by defining some essential notions and formally stating the problem
considered here, then review previous research addressing this problem.

\subsection{Event Logs and Intrusion Detection – Formal Definitions}

Consider a computer network, which we define here as a heterogeneous set of
entities (e.g. users, hosts, files).
Our interest lies in monitoring activity inside this network
in order to detect malicious behavior (defined in more detail below).
A wide variety of data sources can be used, mostly decomposing
into two categories: host logs (e.g. Windows event logs) 
and network logs (e.g. NetFlows).
In order to make our work applicable independently of the specific data source
at hand, we use an abstract definition built upon the notion of event.
An event is generically defined as an interaction between two or more entities,
further characterized by a timestamp, an event type and some optional
additional information.
Authentications can for instance be represented as interactions between users
and hosts, with additional information such as the authentication package used.
Note that this representation is not unique: one could for example treat the
authentication package as an involved entity rather than additional
information, or include the source and destination hosts as two separate
entities.

There are three important aspects in the above definition.
First, events are combinatorial data, in that they are primarily defined as a
combination of entities chosen from finite sets.
But there are several event types, and the observed combinations of entities
depend on the type: for instance, an authentication and a process creation
involve different kinds of entities.
In other words, event logs are intrinsically heterogeneous data.
Finally, each event is associated with a specific timestamp, and the
probability of observing a given event is not necessarily constant in time.
This endows event logs with a temporal dimension.
Although none of these three characteristics is new to the seasoned
statistician, observing all three of them simultaneously is less frequent.

Given a sequence of events, intrusion detection consists in finding a
subset of this sequence reflecting malicious activity.
Here, we consider a continuous monitoring setting: we first use a set of
supposedly benign historical events to infer the normal behavior of the system.
We then observe a stream of new events and, using these events, we try to
detect malicious activity as soon as possible when it happens.
More specifically, we focus on post-compromise detection: we look for an
attacker having already gained a foothold inside the network, trying to
elevate privileges and move laterally while avoiding detection.
This scenario encompasses multiple kinds of threats, from malicious insiders
to so-called Advanced Persistent Threats (APT).
We make two assumptions about the set of malicious events: first, it is
supposed to be small compared to the global volume of observed events, which
is consistent with our threat model (the attacker tries to avoid detection).
Secondly, malicious events are supposed to be distinguishable from benign ones,
meaning that malicious behaviors stand out in a certain way – if they did not,
trying to detect them would be pointless.
These two assumptions justify resorting to statistical anomaly detection.

\subsection{Related Work}

One of the main challenges in event log-based intrusion detection lies in the
peculiar complexity of the data: no off-the-shelf anomaly detection algorithm
simultaneously addresses the three dimensions introduced earlier
(combinatorial, heterogeneous and temporal).
Therefore, bridging the semantic gap between the data and existing models is
the first step of all research addressing this topic.

Most published work achieves this primarily by transforming the data, making
them simpler before feeding them to a standard anomaly detection algorithm.
More specifically, this process often implies aggregating events by entity and
time period, then abstracting the obtained event sets into simple mathematical
objects such as fixed-sized vectors, discrete sequences or graphs.
The intuition behind this aggregation-based approach is that the set of events
involving a given entity summarizes this entity's behavior.
Therefore, applying an anomaly detection algorithm to the aforementioned
mathematical objects amounts to looking for entities which behave unusually
during certain time windows.
Events are typically aggregated by user~\cite{eldardiry2013multi,%
legg2015automated,kent2015authentication,rashid2016new,turcotte2016modelling,%
hu2017anomalous,tuor2017deep,liu2019log2vec,%
powell2020detecting}, but hosts~\cite{yen2013beehive,gonccalves2015big,%
sexton2015attack,veeramachaneni2016ai,bohara2017unsupervised,%
siddiqui2019detecting} or
user-host pairs~\cite{shashanka2016user,adilova2019system} are other possible
aggregation keys.

Although this approach can effectively detect some specific kinds of malicious
behavior, it is intrinsically limited by
the information loss caused by the initial transformation of the data.
In particular, aggregating events by entity hinders full use of
their combinatorial nature: when considering events only from the point
of view of one entity at a time, one cannot fully exploit the patterns of
interaction between entities.
Take for instance a compromised account being used for malicious lateral
movement.
Using an aggregation-based approach, such an account could be flagged as
anomalous for authenticating on an unusual host, or on an unusually large
number of new hosts.
But this criterion is prone to false positives: users frequently visit new
hosts for legitimate reasons.
Further analysis is thus required: for instance, do similar users usually
visit the incriminated hosts?
If so, there may not be sufficient evidence to raise an alert.

A number of recent contributions have thus explored a more relationship-aware
approach.
A typical example is the analysis of user-resource access
matrices~\cite{turcotte2016poisson,tang2017reducing,gutflaish2019temporal}
(or, equivalently, access graphs~\cite{bowman2020detecting,passino2020graph}),
which aims to model the probability of any given user accessing any given
resource (e.g. servers or database tables) based on global access history.
Practical implementations of this idea often borrow from the field of
recommender systems.
User-resource interaction modelling (as well as the closely related concept
of host-host communication modelling~\cite{lee2019anomaly}) effectively
overcomes the main weakness of aggregation-based methods, i.e. the inability
to consider the patterns of association between entities in a global fashion.
However, it remains limited by its dyadic nature: by representing events as
interactions between pairs of entities, it cannot efficiently encode such
simple notions as a remote authentication, which involves at least three
entities (user, source and destination).

Several ideas have been proposed to address this limitation.
Focusing on the remote authentications mentioned above, Siadati and
Memon~\cite{siadati2017detecting} rely on explicit categorical attributes of
users and hosts to characterize authentications (defined as
user-source-destination triples).
They then design a pattern mining methodology to detect unusual combinations
of attributes, in turn marking the corresponding authentications as anomalous.
Tuor et al.~\cite{tuor2018recurrent} introduce a language modelling approach
for event logs, using recurrent neural networks to learn the structure of
normal log lines.
More recently, Leichtnam et al.~\cite{leichtnam2020sec2graph} proposed a
heterogeneous graph representation for various kinds of network events,
detecting anomalous edges in this graph using an autoencoder.

Finally, a very promising approach has been introduced by Amin et
al.~\cite{amin2019cadence}.
Their algorithm, called \textsc{Cadence}, treats event logs as categorical
data.
In other words, each event is defined as a tuple of fixed size, with each
element of the tuple chosen from a distinct and finite set.
\textsc{Cadence} then models the probability distribution of events through
a representation learning approach.
More specifically, this distribution is decomposed into a product of
conditional distributions: instead of directly estimating the joint probability
of a tuple $(x_1,\ldots,x_k)$, \textsc{Cadence} separately models each of the
conditional probabilities
$p(x_2\mid{x}_1)$, $p(x_3\mid{x}_1,x_2)$, up until
$p(x_k\mid{x}_1,\ldots,x_{k-1})$.
The product of these conditional probabilities can then be used as an anomaly
score for the event.
This idea is interesting for several reasons.
First, in contrast
with~\cite{siadati2017detecting,leichtnam2020sec2graph}, it does not rely on
explicit attributes of the entities, such as users' organizational roles.
Secondly, it makes better use of the structure of events
than~\cite{tuor2018recurrent}, leading to a simpler and thus more reliable
model.
Finally, its predictions are built through an understandable and interpretable
reasoning: coming back to the remote authentication example, \textsc{Cadence}
first asks, given the user, how likely it is to observe this user
authenticating from the source host.
Then, given the user and the source, it asks how likely the user is to
authenticate from the source to the destination.
This can be seen as mimicking the way a human analyst would interpret the
event.

Despite these upsides, \textsc{Cadence} also has a few shortcomings.
In particular, it can only handle a single event type, and once the model
has been learned, it does not evolve to adapt to new observations: the only
way to update it is a complete retraining, which is not efficient.
Therefore, we take inspiration from \textsc{Cadence} in building our model,
redesigning it to address these limitations.

\section{Multi-Task Learning for Heterogeneous Higher-Order Interactions}
\label{sec:model}

We now describe the main elements of \textsc{Decades}, focusing on its
combinatorial and heterogeneous aspects.
First of all, Section~\ref{sec:model:representation} formally defines the
mathematical representation of events used throughout this work, along with
some necessary notations.
We then present our statistical model in Section~\ref{sec:model:model}
and explain how its parameters are learned in
Sections~\ref{sec:model:inference} and~\ref{sec:model:mtl}.
Finally, Section~\ref{sec:model:detection} addresses computation of anomaly
scores using the trained model.

\subsection{Representing Events as Higher-Order Interactions}
\label{sec:model:representation}

\setlength{\tabcolsep}{4pt}
\begin{table}[t]
	\centering
	\caption{Key notations}
	\begin{tabular}{lp{.37\textwidth}}
	\toprule
		\textbf{Notation}       & \textbf{Meaning}                               \\
	\midrule
		$\mathcal{V}_k^T$       & Set of all entities of type $k$
                              observed up to time step $T$ ($T$ omitted
                              when irrelevant)                               \\
		$M$                     & Number of entity types                         \\
		$N_e$                   & Number of involved entities for event type $e$ \\
		$\omega=\omega_{1:N_e}$ & Set of entities involved in an event           \\
		$\tau_i^e$              & Type of the $i$-th entity of a type $e$ event  \\
		$\mathbf{x}_v^T$        & Embedding of entity $v$ at time step $T$ ($T$
                              omitted when irrelevant)                       \\
		$w_{ij}^e$              & Weight of the $i$-th entity in predicting the
                              $j$-th entity for event type $e$               \\
		$\boldsymbol\beta_e$    & Latent weight vector specific to event type $e$\\
		$Q$                     & Noise distribution used for NCE                \\
		$K$                     & Number of negative samples used for NCE        \\
		$\sigma_{e,i}$          & Uncertainty in the prediction of the $i$-th
                              entity for a type $e$ event                    \\
	\bottomrule
	\end{tabular}
	\label{tab:notations}
\end{table}

As stated above, the events we consider can be generally defined as
timestamped interactions between entities, further characterized by an
event type and some optional additional information.
In this work, we focus on the first three of these elements: each event is
represented as a triple $(t,e,\omega)$, with $t$ the timestamp, $e$ the
event type and $\omega$ the set of involved entities.

Moreover, each entity also has a type (e.g. user or computer).
Let $M$ denote the number of entity types,
and $\mathcal{V}_k$ the set of all entities of type $k$.
For each event type $e$, we assume that each event of type $e$
involves a fixed number of entities $N_e$.
In addition, these entities have a fixed ordering, such that the set of
involved entities $\omega$ can be uniquely described as
$\omega=\left(\omega_1,\ldots,\omega_{N_e}\right)$, which we abbreviate as
$\omega_{1:N_e}$.
Finally, the types of the entities involved in a type $e$ event are assumed
to be fixed, meaning that the $i$-th entity of a type $e$ event must be of
type $\tau_i^e$.
See Table~\ref{tab:notations} for a summary of the notations used throughout
the paper, and Section~\ref{sec:experiments:dataset} for a concrete
implementation of this framework.

\subsection{Modelling the Conditional Distribution of Events}
\label{sec:model:model}

\begin{figure}[t]
	\centering
	\includegraphics[width=\columnwidth]{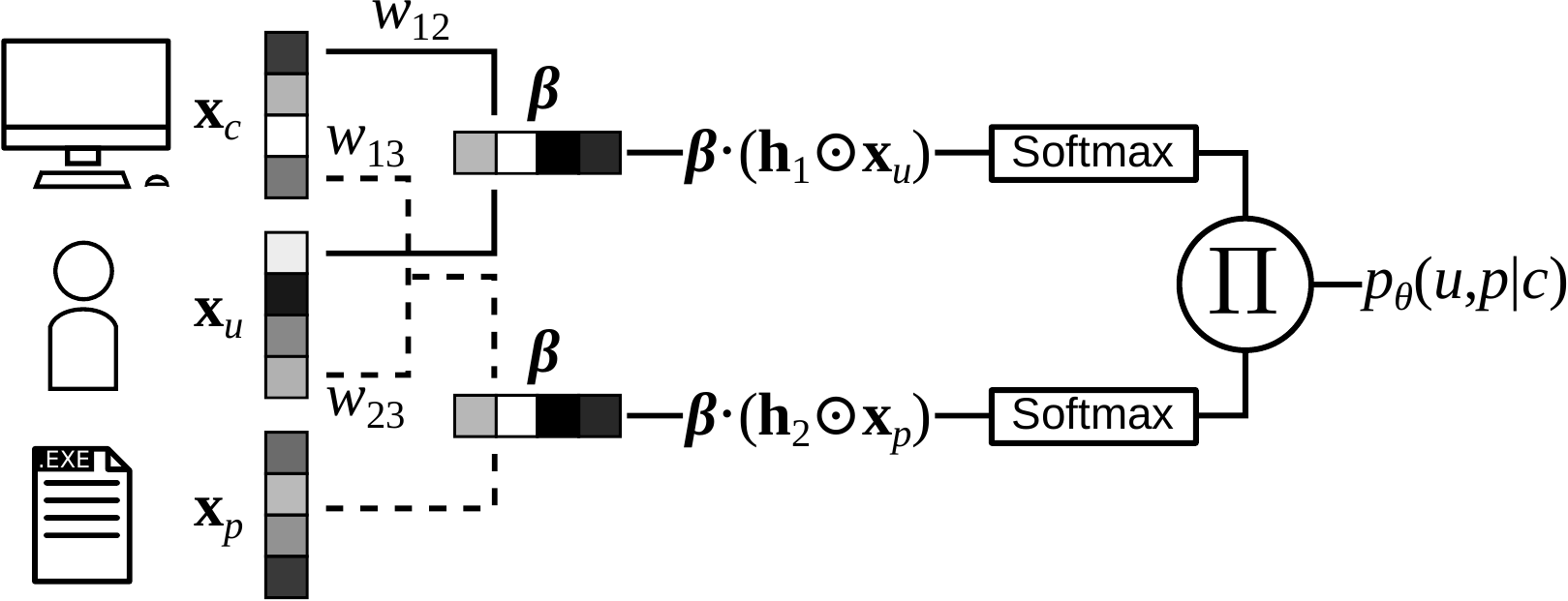}
	\caption{Estimation of the probability of an event
		(here, a process creation).
		For each involved entity $v\in\{u,p\}$,
		its conditional probability depends on
		the weighted dot product between its embedding $\mathbf{x}_v$ and the
		linear combination $\mathbf{h}_i$ of the previous entities' embeddings.
		The product of the conditional probabilities then gives the probability
		of the event given the first entity (here, the host $c$).
    }
	\label{fig:model}
\end{figure}

Following \textsc{Cadence}, we seek to detect anomalous events by modelling
the probability of the involved entities given the event type
and the first entity,
\[
	p(\omega_2,\ldots,\omega_{N_e}\mid \omega_1,e)
	= \prod_{i=2}^{N_e}p(\omega_i\mid \omega_{1:i-1},e).
\]

Each of the conditional probabilities is modelled through representation
learning and softmax regression, as illustrated in Figure~\ref{fig:model}.
We assign to each entity $v$ an embedding $\mathbf{x}_v\in\mathbb{R}^{d}$,
and define the estimated conditional probability
\begin{equation}
    p_{\theta}(\omega_i\mid \omega_{1:i-1},e) =
    \frac{
        \kappa_{e,i}(\omega_i\mid \omega_{1:i-1};\theta)
    }{
        \sum_{v\in\mathcal{V}_{\tau_i^e}}
        \kappa_{e,i}(v\mid \omega_{1:i-1};\theta)
    }.
    \label{eq:prediction}
\end{equation}
This estimated probability depends on the affinity function \[
	\kappa_{e,i}(v\mid \omega_{1:i-1};\theta) =
		\exp\left(\boldsymbol\beta_e\cdot
		(\mathbf{x}_{v}\odot\mathbf{h}_{e,i})\right), \qquad
	\mathbf{h}_{e,i} = \sum_{j=1}^{i-1}w_{ji}^e\mathbf{x}_{\omega_j},
\] where the weights $\{w_{ji}^e\}_{1\leq{j}<i}$ and the latent factor
$\boldsymbol\beta_e$ are parameters specific to the $e$-th event type, and
$\odot$ denotes the element-wise product of two vectors.
We write $\theta$ for the set of parameters of the model, including the
entities' embeddings $\mathbf{x}_v$, the latent factors $\beta_e$
and the weights $w_{ij}^e$ used to compute the $\mathbf{h}_{e,i}$ vectors.

Intuitively, the affinity function $\kappa_{e,i}$ quantifies the compatibility
of a given entity $v$ with the others: if $v$'s embedding $\mathbf{x}_v$ is
similar to the embeddings $\mathbf{x}_{\omega_{1:i-1}}$ of previous entities
$\omega_{1:i-1}$, then the affinity $\kappa_{e,i}(v\mid\omega_{1:i-1};\theta)$
is high, meaning that $v$ is likely to interact with $\omega_{1:i-1}$.
The event type-specific parameters
$\left\{w_{ij}^e\right\}_{1\leq{i}<j\leq{N_e}}$ and $\boldsymbol\beta_e$
allow the model to adapt to the specificities of the different event types:
the latent factor $\boldsymbol\beta_e$ can be seen as a way to select some
dimensions of the embedding space, so that two entities can be likely to
appear together in an event of one specific type and not the others.
In addition, the weights $w_{ij}^e$ are used to adjust the
importance of each entity in predicting other parts of the event.
For instance, when predicting the destination host of an authentication, one
may want to give more or less importance to the user credentials being used or
to the source host.
Here, these weights are tuned automatically when learning the model, which is
also a way to make the latter more interpretable (see
Figure~\ref{fig:weights} for an example).

\subsection{Fitting the Model with Noise Contrastive Estimation}
\label{sec:model:inference}

Having defined the model, we now need to design a training procedure in order
to learn the set of parameters $\theta$ from historical data.
Given an observed event sequence
$\mathcal{O}=\{(t^k,e^k,\omega^k)\}_{k=1}^{n}$, the model is classically
trained by maximizing the conditional log-likelihood \[
	\mathcal{L}(\theta;\mathcal{O}) = \sum_{k=1}^n
		\log p_{\theta}\left(\omega^k_{\geq{2}}\mid e^k,\omega^k_1\right),
\]
or, equivalently, finding a set of parameters $\theta^*$ that minimizes the
negative conditional log-likelihood, translating to
\[\theta^* \in\operatorname*{arg\,min}_{\theta}
-\mathcal{L}(\theta;\mathcal{O}).\]
The traditional machine learning approach finds an approximate solution
to this optimization problem through stochastic gradient descent (SGD),
minimizing the expected value of an event-wise loss function over the training
set $\mathcal{O}$.
Here, the loss function for an event $(t^k,e^k,\omega^k)$ is simply the sum of
the negative conditional log-probabilities of the involved entities, \[
	-\log p_{\theta}\left(\omega^k_{\geq{2}}\mid e^k,\omega^k_1\right)
	= -\sum_{i\geq{2}}\log p_{\theta}(\omega^k_i\mid \omega^k_{1:i-1},e^k).
\]
However, computing each of these conditional probabilities implies summing over
the whole set of entities of a given type (see Equation~\ref{eq:prediction}),
making direct optimization of the loss function too costly.
Following \textsc{Cadence}, we circumvent this issue by using Noise
Contrastive Estimation (NCE~\cite{gutmann2010noise}).

In short, the idea of NCE is to recast the estimation of a probability
distribution as a binary classification problem.
Instead of directly maximizing the probability of each entity $\omega^k_i$,
NCE tries to distinguish it from so-called negative samples, i.e. random
entities sampled from $\mathcal{V}_{\tau_i^e}$ using a known noise
distribution~$Q$.
This leads to a surrogate per-entity loss
\begin{equation*}
\begin{split}
	\ell_i(e^k,\omega^k;\theta) =&
	-\log\mathbb{P}\left[
		\omega^k_i\in\mathcal{D}\mid\omega^k_{1:i-1},e^k;\theta
	\right] \\ &
	- \sum_{j=1}^K
	\log\mathbb{P}\left[
		\tilde{\omega}^k_{i,j}\notin\mathcal{D}\mid\omega^k_{1:i-1},e^k;\theta
	\right],
\end{split}
\end{equation*}
where $K$ denotes the number of negative samples and
$\tilde{\omega}^k_{i,j}\in\mathcal{V}_{\tau_i^e}$ is the $j$-th negative
sample.
The expression \(\mathbb{P}\left[
	v\in\mathcal{D}\mid\omega^k_{1:i-1},e^k;\theta
\right]\) denotes the probability of entity $v$ coming from the true
conditional distribution given $\omega^k_{1:i-1}$ and $e^k$.
In other words, it is the probability of $v$ not being a negative sample,
equal to
\begin{equation*}
\begin{split}
	\mathbb{P}\left[v\in\mathcal{D}\mid\omega^k_{1:i-1},e^k;\theta\right]
	&= \frac{p_{\theta}\left(v\mid e^k,\omega^k_{1:i-1}\right)}
		{p_{\theta}\left(v\mid e^k,\omega^k_{1:i-1}\right)
			+ KQ\left(v\right)} \\
	&\approx \frac{\kappa_{e^k,i}(v\mid \omega^k_{1:i-1};\theta)}
		{\kappa_{e^k,i}(v\mid \omega^k_{1:i-1};\theta)
			+ KQ\left(v\right)}.
\end{split}
\end{equation*}
The approximation on the second line amounts to omitting the normalization
factor – this is what actually makes training computationally efficient.
This approximation was introduced for word embeddings by Mnih and
Teh~\cite{mnih2012fast}, who empirically found out that it did not affect the
performance of the model.
Note that this only works because of the negative sampling scheme induced by
NCE: intuitively, instead of making the true entity as likely as possible, we
try to make it more likely than the negative samples.
Since the normalization factor is the same for the positive and negative
samples, its value has no impact on their ordering by probability.
The NCE loss also depends on the converse probability
$\mathbb{P}\left[v\notin\mathcal{D}\mid\omega^k_{1:i-1},e^k;\theta\right]$,
which is simply obtained as \[
  \mathbb{P}\left[v\notin\mathcal{D}\mid\omega^k_{1:i-1},e^k;\theta\right]=
	1-\mathbb{P}\left[v\in\mathcal{D}\mid\omega^k_{1:i-1},e^k;\theta\right].
\]
Finally, the event-wise loss is
\begin{equation}
	\ell(e^k,\omega^k;\theta) =
		\sum_{i\geq 2}\ell_i(e^k,\omega^k;\theta).
\label{eq:event_loss}
\end{equation}

In order to efficiently learn the true probability distribution, a suitable
noise distribution $Q$ must be designed.
\textsc{Cadence} uses the unigram distribution for each field, meaning that
$Q(v)$ is proportional to the number of occurrences of $v$ in the training
set, denoted $C(v)$.
This is a common choice for learning word embeddings~\cite{mnih2013learning,%
mikolov2013distributed}.
However, we empirically found that using logarithmic unigram counts, i.e.
$Q(v)\propto\log(1+C(v))$, gives better results.
This could be explained by the strong unbalancedness of the unigram
distribution, which leads a significant proportion of the entities to almost
never appear as negative samples, in turn degrading the quality of the
predictions for these entities.
Note that $N_e-1$ distinct log-unigram noise distributions are built for each
event type $e$ (one for each predicted entity).

\subsection{Learning from Heterogeneous Events with Multi-Task Learning}
\label{sec:model:mtl}

NCE allows us to approximate the gradient of the negative log-likelihood for
a batch of events of one given type.
However, one specificity of \textsc{Decades} is that it handles multiple event
types using some shared parameters, namely the embeddings of the
entities.
Hence an additional challenge: when optimizing parameters based on
several (possibly conflicting) objectives, there is no \textit{a priori}
optimal way to balance the respective influences of these objectives in each
parameter update.
For instance, modifying a user $u$'s embedding may increase the predicted
probability of $u$'s authentications and decrease that of the process
creations involving $u$.
How can we find the right balance between these two objectives?

This is a classical multi-task learning (MTL~\cite{caruana1997multitask})
problem.
MTL can be intuitively defined as simultaneously
learning to perform several related tasks, using some shared parameters.
It has given interesting results across various kinds of applications, with
models trained using MTL performing better on each single task than a model
trained to perform this task only.
See~\cite{ruder2017overview} for an introduction to the field and an overview
of existing methods.
Following the terminology proposed therein, our approach can be
described as hard parameter sharing: we perform various prediction tasks
using the same entity embeddings, along with some task specific parameters.
Thus we need a sensible aggregation method for the losses associated with the
tasks, which outputs a single loss that can then be used to perform gradient
descent.

The approach we choose here is the uncertainty-based weighting scheme of
Kendall et al.~\cite{kendall2018multi}.
For each event type $e$ and index $i$, let $\sigma_{e,i}$ denote the
uncertainty associated with predicting the $i$-th entity involved in a type
$e$ event.
We then transform the event-wise loss from Equation~\ref{eq:event_loss} into
\begin{equation}
	\ell_{\text{MT}}(e^k,\omega^k;\theta) = 
	\sum_{i\geq 2} \left\{
		\frac{1}{\sigma_{e^k,i}^2}\ell_i\left(e^k,\omega^k;\theta\right)
		+ \log\sigma_{e^k,i}
	\right\},
\label{eq:mtl_loss}
\end{equation}
and we treat the uncertainties as parameters that are learned along with the
others.
We choose this method both because of its simplicity and its sound theoretical
justification (see~\cite{kendall2018multi} for the detailed theoretical
framework and derivation of the weighting scheme).
As evidenced in Section~\ref{sec:experiments}, this MTL approach allows
\textsc{Decades} to efficiently learn the behavior of entities across different
event types, in turn yielding better performance than similar models trained
for a single type.

Putting it all together, the global loss for the observed sequence
$\mathcal{O}$ is defined as
$J(\theta;\mathcal{O})=\sum_{k=1}^{n}\ell_{\text{MT}}(e^k,\omega^k;\theta)$.
It is minimized using the Adam optimizer~\cite{kingma2014adam}.

\subsection{Using the Learned Model to Compute Anomaly Scores}
\label{sec:model:detection}

Once the parameter set $\hat{\theta}$ has been estimated, it can be used
to assess the degree of abnormality of a given event.
Intuitively, anomalous events should be characterized by a low predicted
probability.
However, the complex nature of the data considered here calls for a more
thorough definition.

First, the predicted probability of an event is in fact a product of
conditional predictions, which do not necessarily behave identically.
In particular, the entities involved in an event are chosen from several
distinct sets (e.g. users, computers).
Since these sets can be of different sizes, the corresponding probability
distributions are dissimilar.
As an illustration, consider sampling uniformly at random from a finite set
$\mathcal{S}$: the probability of each element of $\mathcal{S}$ becomes
smaller as the size of $\mathcal{S}$ increases, but it does not make these
elements anomalous.
Therefore, a proper way to aggregate entity-wise predictions into a global
anomaly score for an event $(t,e,\omega)$ is to use discrete $p$-values,
defined as \[
	p_{e,i} = \sum_{v\in\mathcal{V}_{\tau_i^e}}p_{\hat{\theta}}\left(
		v \mid \omega_{1:i-1},e
	\right)\mathbbm{1}_{\left\{
		p_{\hat{\theta}}\left(
			v \mid \omega_{1:i-1},e
		\right) \leq p_{\hat{\theta}}\left(
			\omega_i \mid \omega_{1:i-1},e
		\right)
	\right\}},
\] where $\mathbbm{1}_{\{\cdot\}}$ denotes the indicator function of a boolean.
A low $p_{e,i}$ means that entity $i$ is not only unlikely given the context,
but also less likely than most other candidate entities.
These $p$-values are then aggregated into a single score, \[
    y = -\frac{1}{N_e-1} \sum_{i=2}^{N_e} \log p_{e,i}.
\]

The second challenge comes from the heterogeneity of events.
Indeed, the model can be better at predicting certain event types than others,
which causes the distribution of anomaly scores to vary across types.
We account for this effect by standardizing the scores independently for each
event type: the mean and standard deviation of anomaly scores for each event
type are evaluated on the training set at the end of training, and the
scores of subsequent events are standardized using these
estimates.

\section{Dealing with Nonstationary Data Streams: Temporal Evolution of the
Model}
\label{sec:evolution}

So far, we have only discussed the initial training of the model.
However, one of the main characteristics of event logs, further discussed in
Section~\ref{sec:evolution:events}, is that they reflect temporally evolving
phenomena.
In Section~\ref{sec:evolution:procedure}, we explain
how \textsc{Decades} adapts to the two main sources of unstability, namely new
entities appearing and existing entities changing their behavior.
Section~\ref{sec:evolution:wrap_up} then wraps up the whole event stream
analysis procedure, including anomaly detection and retraining.

\subsection{On the Temporal Aspect of Event Logs}
\label{sec:evolution:events}

Event logs are records of activity occurring inside a computer network, and
this activity is intrinsically nonstationary.
Therefore, the probability distribution of event logs, which \textsc{Decades}
aims to model, evolves in time as well.
This evolution can be attributed to two distinct factors.
First, new entities appear in the network:
user accounts are created for new employees, new servers are set up, new
software is deployed, and so on.
It is obviously challenging to assess the normality of these entities'
activity without any historical baseline, which is commonly referred to as the
cold start problem.
Secondly, existing entities can also change their behavior for legitimate
reasons: for instance, users can start working on new projects, and servers
can start hosting additional applications.

Such temporal evolution can lead to decaying detection performance unless the
model adapts to it.
However, a reasonable trade-off must be found between fitting the model to the
latest observations and remembering older data.
We propose a tunable procedure to achieve this trade-off,
dealing with both previously unseen entities and evolving behaviors of known
entities.
Since change in the observed data is assumed to reflect a shift in the
entities' activities, the only parameters we update are the embeddings of the
entities.
Thus we replace each embedding $\mathbf{x}_v$ with a sequence
$\mathbf{x}_v^0,\ldots,\mathbf{x}_v^T$, where $\mathbf{x}_v^0$ is the initial
embedding learned during the training phase.
Each time step $\tau\in\{0,\ldots,T\}$ corresponds to the beginning of a
fixed-length window, for instance a day.

\subsection{Adapting to New Data – Updating Procedure}
\label{sec:evolution:procedure}

Let $\mathcal{O}_{T+1}=\{(t^k,e^k,\omega^k)\}_{T\leq{t^k}<T+1}$ denote the set
of events observed between $T$ and $T+1$ (where $t=0$ is the end of the
training phase and the beginning of the detection phase).
We begin with the first identified cause of temporal evolution, namely the
apparition of new entities.
Letting $\mathcal{V}_r^T$ denote the set of entities of type $r$ seen up to
time step $T$, we write
$\tilde{\mathcal{V}}_r^{T+1}=\mathcal{V}_r^{T+1}\setminus\mathcal{V}_r^{T}$
for the set of previously unseen type $r$ entities in $\mathcal{O}_{T+1}$.
For each entity type $r$, we initialize the embedding of each new entity
$v\in\tilde{\mathcal{V}}_r^{T+1}$ as the mean embedding for type $r$, \[
	\mathbf{x}_v^T = \frac{1}{\left|\mathcal{V}_r^{T}\right|}
		\sum_{u\in\mathcal{V}_r^{T}}\mathbf{x}_u^T,
\]
and we use this initial embedding to compute the anomaly scores of events
involving $v$ between $T$ and $T+1$.
Intuitively, in the absence of former activity, a new entity is assumed to
behave similarly to its peers.

All entity embeddings are then tuned in a retraining phase: we optimize the
same loss function as in the training phase (defined in
Equation~\ref{eq:mtl_loss}),
adding regularization terms to prevent overfitting.
This yields a retraining loss
\begin{equation}
\begin{split}
J_{T+1}(\mathcal{O}_{T+1};\theta^{T+1}) =&
    \sum_{T\leq{t^k}<T+1}\ell_{\text{MT}}(e^k,\omega^k) \\
		&+ \lambda_0 R_{\text{new}}(T+1) + \lambda_1 R_{\text{old}}(T+1),
\end{split}
\label{eq:retrain_loss}
\end{equation}
with $\lambda_0,\lambda_1>0$ two hyperparameters, $\theta^{T+1}$ the
parameters at time $T+1$, and
\[
	R_{\text{new}}(T+1) = \sum_{v\in\tilde{\mathcal{V}}^{T+1}}
		\left\|\mathbf{x}_v^{T+1}-\mathbf{x}_v^{T}\right\|_2^2,
\]\[
	R_{\text{old}}(T+1) = \sum_{v\in\mathcal{V}^T}
		\left\|\mathbf{x}_v^{T+1}-\mathbf{x}_v^{T}\right\|_2^2,
\]
where
$\tilde{\mathcal{V}}^{T+1}=\bigcup_{r=1}^{M}\tilde{\mathcal{V}}_r^{T+1}$, and
$\mathcal{V}^T=\bigcup_{r=1}^{M}\mathcal{V}_r^T$.
The intuition behind regularization is that the retraining dataset only covers
a short time span, thus it is not guaranteed to be representative of expected
future behavior.
This uncertainty is alleviated by making the new embeddings close to the
previous ones, which encode some prior knowledge: for already known entities,
the embeddings at time $T$ have been previously trained on former events, hence
they reflect long-term behavior.
As for new entities, their embeddings at time $T$ have been initialized as the
mean embedding of their peers, encoding the idea that they are supposed to
behave according to some global patterns specific to the considered network.
The scaling hyperparameters $\lambda_0$ and $\lambda_1$ control the balance
between prior knowledge and new data: the bigger they are, the less trust is
put by the model into the latest events.
Note that this is equivalent to maximum a posteriori (MAP) estimation of the
embeddings with an independent spherical Gaussian prior placed on each
embedding.
The prior mean is the mean embedding of the corresponding type for new
entities, and the embedding at the previous time step for old entities.
As for the prior variance, it is controlled through $\lambda_0$ and
$\lambda_1$.

Similarly to the training phase, the Adam optimizer is used to minimize the
loss function from Equation~\ref{eq:retrain_loss}.
We iterate on the retraining set $\mathcal{O}_{T+1}$ as long as the validation
loss on a small held-out part of $\mathcal{O}_{T+1}$ significantly decreases.

\subsection{Wrap-Up – Operating the Model for Intrusion Detection}
\label{sec:evolution:wrap_up}

\begin{algorithm}[t]
	\KwData{Event stream $\{(t^k,e^k,\omega^k)\}_{k\geq{0}}$, learned parameters
 		$\theta^0$}
	\KwResult{Sequence of anomaly scores $\{y^k\}_{k\geq{0}}$}
	\For{time step $T\geq{0}$}{
		$\mathcal{O}_{T+1}\leftarrow\{(t^k,e^k,\omega^k)\}_{T\leq{t^k}<T+1}$\;
		\tcc{analyze incoming events}
        \For{
            $(t^k,e^k,\omega^k)\in\mathcal{O}_{T+1}$}{
			\tcc{look for new entities}
      \For{
        $v\in\omega^k$}{
				\If{$v\notin\mathcal{V}^{T}$}{
					$r\leftarrow\textsc{Type}(v)$\;
					\(\mathbf{x}_v^T\leftarrow\frac{1}{
					\left|\mathcal{V}_r^{T}\right|}
					\sum_{u\in\mathcal{V}_r^{T}}\mathbf{x}_u^T\)\;
				}
			}
			$y^k=\textsc{ComputeAnomalyScore}(\omega^k,e^k;\theta^{T})$\;
		}
		remove malicious events from $\mathcal{O}_{T+1}$\;
		\tcc{update entity embeddings}
		$\theta^{T+1}\leftarrow\theta^{T}$\;
		\Repeat{convergence}{
			\(\theta^{T+1}\leftarrow\textsc{SgdEpoch}(
			\mathcal{O}_{T+1},\theta^{T+1})\)\;
		}
	}
	\caption{Analyzing a stream of events using the trained model.
	}
	\label{alg:decades}
\end{algorithm}

We conclude this section by summarizing how \textsc{Decades} could be deployed
as an intrusion detection system.
First, initial parameters $\theta^0$ are learned from historical data
(Sections~\ref{sec:model:inference} and~\ref{sec:model:mtl}).
Then, for each time period $[T,T+1)$, the collected events are first assigned
an anomaly score using current parameters $\theta^T$
(Section~\ref{sec:model:detection}), then sorted from the most to the least
anomalous.
A security expert investigates the top anomalies, and the model is then updated
to obtain the new parameters $\theta^{T+1}$
(Section~\ref{sec:evolution:procedure}).
Note that malicious events flagged by the expert should be removed from the
retraining set to avoid polluting the model.
This complete operating procedure is summarized in Algorithm~\ref{alg:decades}.

\section{Experiments}
\label{sec:experiments}

We now present our empirical analysis and evaluation of \textsc{Decades}.
Experiments are performed using a real event log dataset, which we describe in
Section~\ref{sec:experiments:dataset}.
We then compare the detection performance of \textsc{Decades} for this dataset
with several state-of-the-art baselines and discuss some properties of the
trained model in Section~\ref{sec:experiments:comparison}.

\subsection{Implementation, Dataset and Performance Metrics}
\label{sec:experiments:dataset}

\setlength{\tabcolsep}{6pt}
\begin{table*}[t]
	\centering
	\caption{Entity types defined in the LANL dataset, along with their
		arities in the training, test and entire datasets.
    }
	\begin{tabular}{lrrrrrr}
	\toprule
		\multirow{2}{*}{\textbf{Name}}
		& \multicolumn{2}{c}{\textbf{Arity (train)}}
		& \multicolumn{2}{c}{\textbf{Arity (test)}}
		& \multicolumn{2}{c}{\textbf{Arity (all)}}		\\
		\noalign{\vspace{1pt}}\cline{2-7}\noalign{\vspace{4pt}}
		& \textbf{Total} & \textbf{Malicious}
		& \textbf{Total} & \textbf{Malicious}
		& \textbf{Total} & \textbf{Malicious}						\\
	\midrule
		User 		& 12\,164 & 7 & 10\,767 & 71 & 13\,176 & 76 \\
		Computer 	& 12\,264 & 27 & 11\,943 & 234 & 13\,090 & 255 \\
		Authentication type 	& 23 & 1 & 21 & 1 & 24 & 1 \\
		Process 	& 1566 & 0 & 1531 & 0 & 1593 & 0 \\
	\bottomrule
	\end{tabular}
	\label{tab:entity_types}
\end{table*}

\setlength{\tabcolsep}{5pt}
\begin{table*}[t]
    \centering
    \caption{Events in the LANL dataset: event types defined for this work,
        ordered types and meaning of the involved entities, and event counts.
        The entity types are user (U), computer (C), authentication type (T)
        and process (P).
    }
    \begin{tabular}{lllrr}
    \toprule
        \multirow{2}{*}{\textbf{Name}}
        & \multirow{2}{*}{\textbf{Entity types}}
        & \multirow{2}{*}{\textbf{Entity meanings}}
        & \multicolumn{2}{c}{\textbf{\#\,Total (\#\,malicious)}}            \\
    \noalign{\vspace{1pt}}\cline{4-5}\noalign{\vspace{4pt}}
        & & & \textbf{Train}    & \textbf{Test}                             \\
        \midrule
        
        \multirow{2}{*}{Local authentication}
        & \multirow{2}{*}{(U, T, C)}
        & Credentials used, authentication type,
        & \multirow{2}{*}{3\,418\,117 (0)}
        & \multirow{2}{*}{2\,173\,349 (0)} \\
        &
        & host where logon happens
        &
        & \\
        
        \midrule

        \multirow{2}{*}{Remote authentication}
        & \multirow{2}{*}{(U, T, C, C)}
        & Credentials used, authentication type,
        & \multirow{2}{*}{13\,198\,597 (50)}
        & \multirow{2}{*}{8\,310\,963 (473)} \\
        &
        & source host, destination host
        &
        & \\
        
        \midrule

        \multirow{2}{*}{Process creation}
        & \multirow{2}{*}{(C, U, P)}
        & Host where the process is created, user
        & \multirow{2}{*}{4\,949\,066 (0)}
        & \multirow{2}{*}{2\,758\,106 (0)} \\
        &
        & creating the process, process name
        &
        & \\
        \bottomrule
    \end{tabular}
    \label{tab:event_types}
\end{table*}

We implemented \textsc{Decades} in Python 3.9, mostly using
PyTorch~\cite{paszke2019pytorch}.
This implementation is openly available online\footnote{
https://github.com/cl-anssi/DECADES}.
Experiments were run on a Debian 10 machine with 128GB RAM and a Tesla V100
GPU.
Hyperparameters were tuned to maximize detection performance, yielding the
following values: embedding dimension $d=64$, number of negative samples per
batch $K=10$, regularization coefficients $\lambda_0=10^{-4}$ and
$\lambda_1=1$.
The model was initially trained for 30 epochs with a learning rate of $10^{-3}$
and 5000-event batches, and retrained after each day of the test
set with a learning rate of $10^{-4}$ and 256-event batches.

All experiments are based on the "Comprehensive, Multi-Source Cyber-Security
Events" dataset~\cite{kent2015cybersecurity,kent2015cyberdata}
released by the Los Alamos National Laboratory.
This dataset contains various types of event logs recorded during 58 days on
a real enterprise network.
A red team exercise took place during this period, and authentication events
associated with this exercise are labelled.
These labels are used to evaluate detection performance.
The dataset is preprocessed for our experiments:
first, only authentication and process-related events are included.
Regarding authentications, we only consider "LogOn" events for which the
destination user is a user account (i.e. not a computer account or built-in
account such as "LOCAL SYSTEM").
We distinguish two kinds of authentications: local ones, each of which involves
three entities (destination user, authentication type, computer), and remote
ones, which involve four entities (destination user, authentication type,
source computer, destination computer).
Note that the authentication type is here defined as the pair (authentication
package, logon type), which is more informative from a security point of view
than either of these fields alone.
As for processes, we only include process creations, excluding process stops as
they seem less relevant from an intrusion detection perspective.
Each of them involves three entities (computer, user, process name).
Process names appearing less than 40 times in the dataset are replaced with
a "rare process" token.
The first 8 days are used for training, and the next 5 days are used for
evaluation.
See Table~\ref{tab:event_types} for a summary of the considered event types,
and Table~\ref{tab:entity_types}
for a list of the entity types and their arities in the dataset.

Two criteria are used to evaluate detection performance.
The first one is the Receiver Operating Characteristic (ROC) curve,
which plots the detection rate on the whole test dataset as a function of
the false positive rate (FPR) for varying detection thresholds.
Due to the considerable volume of real-life event streams,
only the leftmost part of the ROC curve (FPR$\leq$1\%) is considered: with
millions of events generated daily, a 1\% false positive rate already
amounts to tens of thousands of false positives to investigate every day.
Therefore, detection thresholds leading to a greater FPR can be considered
irrelevant.
The second metric is the detection rate for a fixed daily budget $B$ of
investigated events, denoted DR-$B$.
It gives a more realistic view of what could actually be detected when
monitoring activity in an enterprise network.
Note that since only authentications are labelled as benign or malicious in
the LANL dataset, these two criteria are always computed only for
authentication events from the test set.

\subsection{Experimental Results}
\label{sec:experiments:comparison}

\begin{figure}[t]
    \centering
    \includegraphics[width=\columnwidth]{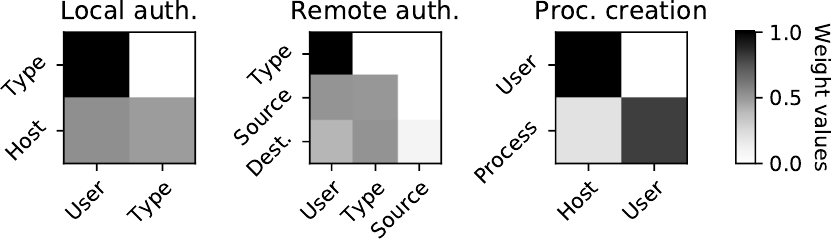}
    \caption{Weights learned by \textsc{Decades} on the LANL dataset.
    Vertical labels stand for entities being predicted, while horizontal
    labels indicate entities being used for the prediction.
    For instance, when predicting the destination of a remote
    authentication, the model relies mostly on the authentication type.
}
    \label{fig:weights}
\end{figure}

\begin{figure}[t]
    \includegraphics[width=.4\textwidth]{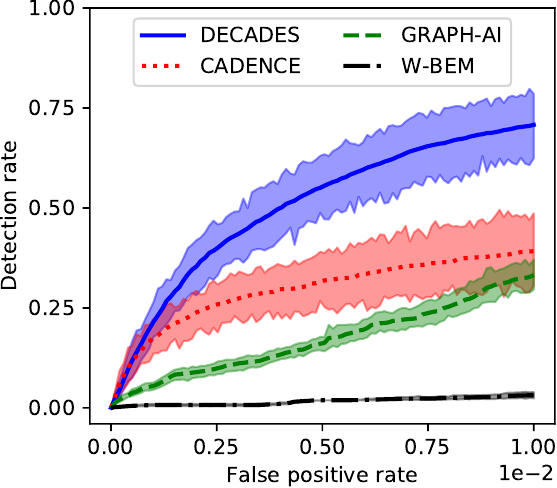}
    \caption{Truncated ROC curves of the 4 evaluated methods on the
    authentication events, with 95\% confidence intervals.}
    \label{fig:comp_roc}
\end{figure}

We now present our experimental study of \textsc{Decades}, addressing three
main questions: first, what does the model actually learn about the dataset?
Secondly, does \textsc{Decades} perform well at detecting malicious behavior?
Finally, to what extent do hyperparameters affect detection performance?

As for the first question, Figure~\ref{fig:weights} displays the weights
$w^e_{ij}$ obtained after training.
One rather unexpected outcome is that the destination of a remote
authentication is mostly predicted through its type, while little importance
is given to the user and source.
One possible explanation is that the authentication type is correlated with the
functional role of the destination host: for instance, most servers should be
authenticated to using Kerberos, but some of them might only support NTLM.
Similarly, remote desktop sessions might be primarily used by help desk
employees to connect to malfunctioning workstations, while servers might be
administered using different tools.

\setlength{\tabcolsep}{4pt}
\begin{table*}[t]
    \centering
    \caption{Mean and 95\% confidence interval of each evaluation metric.
        AUC@1\% is the normalized area under the truncated ROC curve, and
        DR-$B$ is the proportion of red team events detected by extracting the
        $B$ most anomalous events each day.
        The best mean score for each metric is marked in bold.}
    \begin{tabular}{lrrrrr}
        \toprule
            \textbf{Method}
            & \textbf{AUC@1\%} & \textbf{DR-1K} & \textbf{DR-5K}
            & \textbf{DR-10K} & \textbf{DR-20K} \\
        \midrule
            \textsc{Decades} & \textbf{0.497$\pm$0.085} & 0.081$\pm$0.032 & \textbf{0.333$\pm$0.060} & \textbf{0.482$\pm$0.076} & \textbf{0.640$\pm$0.099} \\
            \textsc{Cadence}  & 0.291$\pm$0.063 & \textbf{0.093$\pm$0.034} & 0.218$\pm$0.051 & 0.276$\pm$0.062 & 0.344$\pm$0.086 \\
            \textsc{GraphAI} & 0.170$\pm$0.019 & 0.032$\pm$0.010 & 0.084$\pm$0.014 & 0.116$\pm$0.013 & 0.218$\pm$0.030 \\
            W-BEM & 0.016$\pm$0.002 & 0.004$\pm$0.002 & 0.009$\pm$0.003 & 0.016$\pm$0.002 & 0.025$\pm$0.005 \\
        \bottomrule
    \end{tabular}
    \label{tab:comp_eval}
\end{table*}

\begin{figure*}[t]
  \begin{minipage}{.24\textwidth}
    \centering
    \begin{subfigure}[t]{\textwidth}
      \centering
      \includegraphics[width=\textwidth]{%
        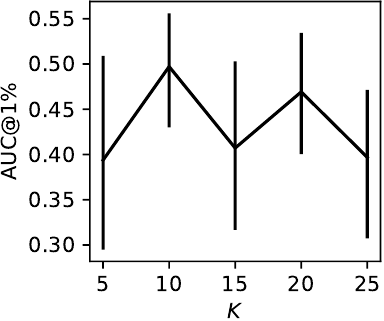}
      \caption{Number of negative samples $K$}
      \label{fig:sensitivity:sensitivity_k}
    \end{subfigure}
  \end{minipage}
  \hfill
  \begin{minipage}{.24\textwidth}
    \centering
    \begin{subfigure}[t]{.982\textwidth}
      \centering
      \includegraphics[width=\textwidth]{%
        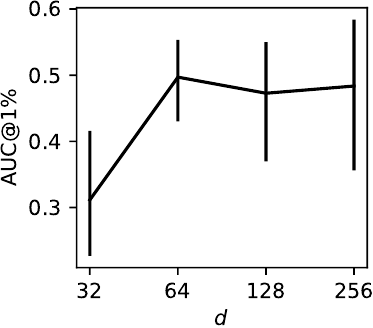}
      \caption{Latent space dimension $d$}
      \label{fig:sensitivity:sensitivity_d}
    \end{subfigure}
  \end{minipage}
  \hfill
  \begin{minipage}{.24\textwidth}
    \centering
    \begin{subfigure}[t]{\textwidth}
      \centering
      \includegraphics[width=\textwidth]{%
        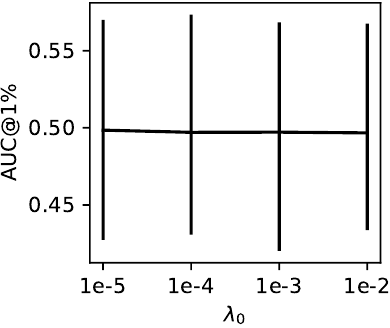}
      \caption{Regularization coefficient $\lambda_0$}
      \label{fig:sensitivity:sensitivity_l0}
    \end{subfigure}
  \end{minipage}
  \hfill
  \begin{minipage}{.24\textwidth}
    \centering
    \begin{subfigure}[t]{\textwidth}
      \centering
      \includegraphics[width=\textwidth]{%
        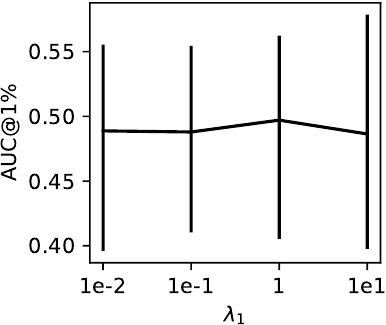}
      \caption{Regularization coefficient $\lambda_1$}
      \label{fig:sensitivity:sensitivity_l1}
    \end{subfigure}
  \end{minipage}

  \captionsetup{width=\textwidth}
  \caption{Area under the truncated ROC curve with 95\% confidence intervals
    for several values of the hyperparameters.}
  \label{fig:sensitivity}
\end{figure*}

In order to answer the second question, we compare \textsc{Decades} with three
previously published methods:
\textsc{Cadence}~\cite{amin2019cadence} and the language model
of~\cite{tuor2018recurrent} (referred to as W-BEM) are closest to our
approach, making them the most relevant references.
In particular, they both aim to fully address the combinatorics of events,
not restricting themselves to pairwise interactions or entity-centric
modelling.
However, they focus only on authentications, whereas jointly modelling
several event types is a key feature of our approach.
We also include the graph-based model of~\cite{bowman2020detecting} (referred
to as \textsc{GraphAI}), which differs by only modelling pairwise user-user and
user-host interactions.
Note that sec2graph~\cite{leichtnam2020sec2graph} and the pattern-based
approach of~\cite{siadati2017detecting} would have been interesting baselines,
but they both rely on explicit attributes of the entities, which are not
available in the LANL dataset.
The open-source implementation of W-BEM was used for the experiments.
We reimplemented \textsc{GraphAI} based on~\cite{bowman2020detecting} and
further precisions obtained from the authors.
Finally, \textsc{Cadence} was implemented based only on \cite{amin2019cadence}
since the authors did not respond to our requests for code.
As for hyperparameters, we reused those provided in the corresponding papers,
except when different values gave better results.

Since all considered algorithms give non-deterministic results (mostly because
of random parameter initialization and stochastic optimization procedures),
each of them is run 20 times.
We then report the mean and 95\% confidence interval (obtained through a
nonparametric bootstrap procedure) for each metric.
Results are displayed in Table~\ref{tab:comp_eval} and
Figure~\ref{fig:comp_roc}.
\textsc{Decades} globally outperforms competing algorithms, with significantly
higher scores for all but one performance metric, namely the detection rate
for a daily investigation budget of 1\,000 events.
Note that these scores remain rather low: for instance, investigating
10\,000 events each day is a considerable amount of work, and it only leads
to a 48\% detection rate using \textsc{Decades}.
However, it should be kept in mind that our aim is only to
sort events so that the most suspicious ones rank close to the top, and further
processing steps should then be applied to obtain more exploitable results (see
Section~\ref{sec:discussion} for more details).

As for competing methods' performance, the best contender is \textsc{Cadence}.
This confirms that \textsc{Cadence}'s approach to modelling the combinatorics
of events, which \textsc{Decades} also relies upon, is indeed the most
effective one.
Next up is \textsc{GraphAI}, which only models dyadic interactions instead of
fully leveraging the combinatorics of events.
This simplified description of events can thus be considered too crude, once
again backing up our hypothesis regarding the importance of factoring in the
complexity of the data.
Finally, W-BEM performs surprisingly poorly, which can be attributed to the
inadequate data specification it relies upon.
Indeed, the idea behind W-BEM is to treat events as sentences, with each
involved entity representing a word.
However, this representation ignores the specific structure of event logs:
the number of involved entities as well as their types are fixed for all events
of a given type, making the actual sample space much smaller than the set of
all sequences of entities.
In other words, W-BEM gives strictly positive probability to many samples
which are actually not events and can thus never happen.
This can be expected to make the model less efficient at learning to predict
actual events.
From a more general perspective, our experiments sustain our initial claim:
carefully adapting the statistical model to the nature of the data leads to a
better intrusion detection system.

We finally address the third question -- namely, how sensitive is
\textsc{Decades}'s detection performance to the value of each hyperparameter?
To that end, we study the evolution of the area under the truncated ROC curve
as one hyperparameter varies while the others are kept fixed to their optimal
values.
The results are displayed in Figure~\ref{fig:sensitivity}.
The most striking observation is that the regularization coefficients
$\lambda_0$ and $\lambda_1$ have no significant influence, suggesting that
retraining makes little difference in our experiments.
This is somewhat understandable given the short duration of the test phase, and
it thus does not refute the importance of model updating in a real network
monitoring setting.
As for the other hyperparameters, the latent space dimension $d$ essentially
controls the number of parameters of the model.
As a consequence, it should be tuned so as to avoid both underfitting and
overfitting.
These two unwanted configurations can be observed in
Figure~\ref{fig:sensitivity:sensitivity_d}, with the lowest value of $d$
yielding significantly inferior results and the highest one leading to
increased uncertainty.
Finally, the number of negative samples $K$ relates to the trade-off between
the quality of the approximation of the log-likelihood gradient provided by NCE
and the computational cost of training.
Figure~\ref{fig:sensitivity:sensitivity_k} shows that increasing $K$
beyond 10 does not make the model better at detecting malicious events.
Note that detection performance even tends to decrease for high values of $K$,
which may result from a kind of overfitting: as $K$ increases, the model more
accurately fits the distribution of the training data.
This might lead it to assign higher anomaly scores to rare but benign events,
in turn yielding more false positives.

\section{Discussion and Future Work}
\label{sec:discussion}

We now discuss some limitations and possible future improvements of our method.
First of all, being an event log-based anomaly detection algorithm,
\textsc{Decades} obviously has some common flaws: it relies on the integrity of
the data, which means that advanced intruders could evade detection by
tampering with the logs, and it actually detects unusual rather than malicious
behavior.
While the former is entirely out of the scope of this work, the latter can lead
to interesting developments.
In particular, the resilience of \textsc{Decades} to deliberate poisoning of
the training set could be studied.

The most important improvement that could be brought to our approach is an
interface making the alerts more exploitable.
Two main directions could be explored.
The first one is the interpretation of each anomalous event, which could rely
on counterfactual explanations: given the entities involved in an anomalous
event, which replacements would make the event normal?
For instance, given the user, type and source host of an anomalous remote
authentication, which possible destinations would be considered normal?
The other way to make the results more usable is alert aggregation and
correlation.
Indeed, \textsc{Decades} reduces the workload of security experts by ranking
events in descending order of anomalousness, allowing the expert to focus on
the most anomalous events first.
However, the number of events to review daily remains significant and needs to
be further reduced, which could be done by merging events resulting from the
same high-level activity.
This issue has been widely addressed in the field of intrusion detection
system alert postprocessing~\cite{valdes2001probabilistic,%
julisch2003clustering,roundy2017smoke,haas2018gac,lin2018collaborative}.

Finally, improving the updating procedure of the model could be a more
theoretical research direction.
In particular, the procedure exposed here does not take into account the
uncertainty associated with the estimated parameters.
This can be problematic when operating on a long time range: by forgetting
every day how confident the model was about each parameter, we might propagate
errors due to unreliable estimates.
A fully Bayesian approach, similar to the one proposed by~\cite{lee2019anomaly}
for host communication graphs, could solve this issue.

\section{Conclusion}
\label{sec:conclusion}

We propose \textsc{Decades}, a novel anomaly detection algorithm for event log
data.
This algorithm extends previous work in order to simultaneously address the
three main challenges associated with event logs: combinatorial observations,
heterogeneity and temporal concept drift.
Our experiments show that handling the full complexity of event log data
through a bespoke statistical model yields better detection results than the
converse approach, namely simplifying the data enough to make them fit into
more traditional anomaly detection frameworks.
Therefore, we argue that this research direction should be explored further
in order to build effective and robust intrusion detection systems for event
logs.

One of the strengths of \textsc{Decades} is its adaptability: by making few
assumptions about the content of the modelled events, we aim to make our
algorithm independent of the chosen logging system.
It would thus be interesting to extend the empirical study to other datasets,
including for instance network logs.
Our method could also be further improved through better alert interpretability
and correlation, as well as uncertainty-aware parameter updates.
With such additional features, the proposed model could evolve into a generic
and reliable intrusion detection algorithm for various kinds of log data.

%
%
%
\balance
\bibliographystyle{ACM-Reference-Format}
\bibliography{references}


\begin{thebibliography}{43}


\ifx \showCODEN    \undefined \def \showCODEN     #1{\unskip}     \fi
\ifx \showDOI      \undefined \def \showDOI       #1{#1}\fi
\ifx \showISBNx    \undefined \def \showISBNx     #1{\unskip}     \fi
\ifx \showISBNxiii \undefined \def \showISBNxiii  #1{\unskip}     \fi
\ifx \showISSN     \undefined \def \showISSN      #1{\unskip}     \fi
\ifx \showLCCN     \undefined \def \showLCCN      #1{\unskip}     \fi
\ifx \shownote     \undefined \def \shownote      #1{#1}          \fi
\ifx \showarticletitle \undefined \def \showarticletitle #1{#1}   \fi
\ifx \showURL      \undefined \def \showURL       {\relax}        \fi
\providecommand\bibfield[2]{#2}
\providecommand\bibinfo[2]{#2}
\providecommand\natexlab[1]{#1}
\providecommand\showeprint[2][]{arXiv:#2}

\bibitem[\protect\citeauthoryear{Adilova, Natious, Chen, Thonnard, and
  Kamp}{Adilova et~al\mbox{.}}{2019}]%
        {adilova2019system}
\bibfield{author}{\bibinfo{person}{Linara Adilova}, \bibinfo{person}{Livin
  Natious}, \bibinfo{person}{Siming Chen}, \bibinfo{person}{Olivier Thonnard},
  {and} \bibinfo{person}{Michael Kamp}.} \bibinfo{year}{2019}\natexlab{}.
\newblock \showarticletitle{System Misuse Detection via Informed Behavior
  Clustering and Modeling}. In \bibinfo{booktitle}{\emph{DSN-W}}.
\newblock


\bibitem[\protect\citeauthoryear{Amin, Garg, and Coskun}{Amin
  et~al\mbox{.}}{2019}]%
        {amin2019cadence}
\bibfield{author}{\bibinfo{person}{Mohammad~Ruhul Amin},
  \bibinfo{person}{Pranav Garg}, {and} \bibinfo{person}{Baris Coskun}.}
  \bibinfo{year}{2019}\natexlab{}.
\newblock \showarticletitle{CADENCE: Conditional Anomaly Detection for Events
  Using Noise-Contrastive Estimation}. In \bibinfo{booktitle}{\emph{AISec}}.
\newblock


\bibitem[\protect\citeauthoryear{Bohara, Noureddine, Fawaz, and Sanders}{Bohara
  et~al\mbox{.}}{2017}]%
        {bohara2017unsupervised}
\bibfield{author}{\bibinfo{person}{Atul Bohara}, \bibinfo{person}{Mohammad~A
  Noureddine}, \bibinfo{person}{Ahmed Fawaz}, {and} \bibinfo{person}{William~H
  Sanders}.} \bibinfo{year}{2017}\natexlab{}.
\newblock \showarticletitle{An Unsupervised Multi-Detector Approach for
  Identifying Malicious Lateral Movement}. In \bibinfo{booktitle}{\emph{SRDS}}.
\newblock


\bibitem[\protect\citeauthoryear{Bowman, Laprade, Ji, and Huang}{Bowman
  et~al\mbox{.}}{2020}]%
        {bowman2020detecting}
\bibfield{author}{\bibinfo{person}{Benjamin Bowman}, \bibinfo{person}{Craig
  Laprade}, \bibinfo{person}{Yuede Ji}, {and} \bibinfo{person}{H~Howie Huang}.}
  \bibinfo{year}{2020}\natexlab{}.
\newblock \showarticletitle{Detecting Lateral Movement in Enterprise Computer
  Networks with Unsupervised Graph {AI}}. In \bibinfo{booktitle}{\emph{RAID}}.
\newblock


\bibitem[\protect\citeauthoryear{Caruana}{Caruana}{1997}]%
        {caruana1997multitask}
\bibfield{author}{\bibinfo{person}{Rich Caruana}.}
  \bibinfo{year}{1997}\natexlab{}.
\newblock \showarticletitle{Multitask learning}.
\newblock \bibinfo{journal}{\emph{Mach. Learn.}} \bibinfo{volume}{28},
  \bibinfo{number}{1} (\bibinfo{year}{1997}), \bibinfo{pages}{41--75}.
\newblock


\bibitem[\protect\citeauthoryear{Eldardiry, Bart, Liu, Hanley, Price, and
  Brdiczka}{Eldardiry et~al\mbox{.}}{2013}]%
        {eldardiry2013multi}
\bibfield{author}{\bibinfo{person}{Hoda Eldardiry}, \bibinfo{person}{Evgeniy
  Bart}, \bibinfo{person}{Juan Liu}, \bibinfo{person}{John Hanley},
  \bibinfo{person}{Bob Price}, {and} \bibinfo{person}{Oliver Brdiczka}.}
  \bibinfo{year}{2013}\natexlab{}.
\newblock \showarticletitle{Multi-domain information fusion for insider threat
  detection}. In \bibinfo{booktitle}{\emph{S\&P Workshops}}.
\newblock


\bibitem[\protect\citeauthoryear{Gon{\c{c}}alves, Bota, and
  Correia}{Gon{\c{c}}alves et~al\mbox{.}}{2015}]%
        {gonccalves2015big}
\bibfield{author}{\bibinfo{person}{Daniel Gon{\c{c}}alves},
  \bibinfo{person}{Jo{\~a}o Bota}, {and} \bibinfo{person}{Miguel Correia}.}
  \bibinfo{year}{2015}\natexlab{}.
\newblock \showarticletitle{Big data analytics for detecting host misbehavior
  in large logs}. In \bibinfo{booktitle}{\emph{Trustcom/BigDataSE/ISPA}}.
\newblock


\bibitem[\protect\citeauthoryear{Gutflaish, Kontorovich, Sabato, Biller, and
  Sofer}{Gutflaish et~al\mbox{.}}{2019}]%
        {gutflaish2019temporal}
\bibfield{author}{\bibinfo{person}{Eyal Gutflaish}, \bibinfo{person}{Aryeh
  Kontorovich}, \bibinfo{person}{Sivan Sabato}, \bibinfo{person}{Ofer Biller},
  {and} \bibinfo{person}{Oded Sofer}.} \bibinfo{year}{2019}\natexlab{}.
\newblock \showarticletitle{Temporal anomaly detection: calibrating the
  surprise}. In \bibinfo{booktitle}{\emph{AAAI}}.
\newblock


\bibitem[\protect\citeauthoryear{Gutmann and Hyv{\"a}rinen}{Gutmann and
  Hyv{\"a}rinen}{2010}]%
        {gutmann2010noise}
\bibfield{author}{\bibinfo{person}{Michael Gutmann} {and} \bibinfo{person}{Aapo
  Hyv{\"a}rinen}.} \bibinfo{year}{2010}\natexlab{}.
\newblock \showarticletitle{Noise-contrastive estimation: A new estimation
  principle for unnormalized statistical models}. In
  \bibinfo{booktitle}{\emph{AISTATS}}.
\newblock


\bibitem[\protect\citeauthoryear{Haas and Fischer}{Haas and Fischer}{2018}]%
        {haas2018gac}
\bibfield{author}{\bibinfo{person}{Steffen Haas} {and} \bibinfo{person}{Mathias
  Fischer}.} \bibinfo{year}{2018}\natexlab{}.
\newblock \showarticletitle{GAC: graph-based alert correlation for the
  detection of distributed multi-step attacks}. In
  \bibinfo{booktitle}{\emph{SAC}}.
\newblock


\bibitem[\protect\citeauthoryear{Hu, Tang, and Lin}{Hu et~al\mbox{.}}{2017}]%
        {hu2017anomalous}
\bibfield{author}{\bibinfo{person}{Qiaona Hu}, \bibinfo{person}{Baoming Tang},
  {and} \bibinfo{person}{Derek Lin}.} \bibinfo{year}{2017}\natexlab{}.
\newblock \showarticletitle{Anomalous User Activity Detection in Enterprise
  Multi-source Logs}. In \bibinfo{booktitle}{\emph{ICDM Workshops}}.
\newblock


\bibitem[\protect\citeauthoryear{Julisch}{Julisch}{2003}]%
        {julisch2003clustering}
\bibfield{author}{\bibinfo{person}{Klaus Julisch}.}
  \bibinfo{year}{2003}\natexlab{}.
\newblock \showarticletitle{Clustering intrusion detection alarms to support
  root cause analysis}.
\newblock \bibinfo{journal}{\emph{TISSEC}} \bibinfo{volume}{6},
  \bibinfo{number}{4} (\bibinfo{year}{2003}), \bibinfo{pages}{443--471}.
\newblock


\bibitem[\protect\citeauthoryear{Kendall, Gal, and Cipolla}{Kendall
  et~al\mbox{.}}{2018}]%
        {kendall2018multi}
\bibfield{author}{\bibinfo{person}{Alex Kendall}, \bibinfo{person}{Yarin Gal},
  {and} \bibinfo{person}{Roberto Cipolla}.} \bibinfo{year}{2018}\natexlab{}.
\newblock \showarticletitle{Multi-task learning using uncertainty to weigh
  losses for scene geometry and semantics}. In
  \bibinfo{booktitle}{\emph{CVPR}}.
\newblock


\bibitem[\protect\citeauthoryear{Kent}{Kent}{2015a}]%
        {kent2015cyberdata}
\bibfield{author}{\bibinfo{person}{Alexander~D. Kent}.}
  \bibinfo{year}{2015}\natexlab{a}.
\newblock \bibinfo{title}{Comprehensive, Multi-Source Cyber-Security Events}.
\newblock \bibinfo{howpublished}{Los Alamos National Laboratory}.
\newblock
\urldef\tempurl%
\url{https://doi.org/10.17021/1179829}
\showDOI{\tempurl}


\bibitem[\protect\citeauthoryear{Kent}{Kent}{2015b}]%
        {kent2015cybersecurity}
\bibfield{author}{\bibinfo{person}{Alexander~D. Kent}.}
  \bibinfo{year}{2015}\natexlab{b}.
\newblock \showarticletitle{Cybersecurity Data Sources for Dynamic Network
  Research}. In \bibinfo{booktitle}{\emph{Dynamic Networks in Cybersecurity}}.
  \bibinfo{publisher}{Imperial College Press}.
\newblock


\bibitem[\protect\citeauthoryear{Kent, Liebrock, and Neil}{Kent
  et~al\mbox{.}}{2015}]%
        {kent2015authentication}
\bibfield{author}{\bibinfo{person}{Alexander~D Kent}, \bibinfo{person}{Lorie~M
  Liebrock}, {and} \bibinfo{person}{Joshua~C Neil}.}
  \bibinfo{year}{2015}\natexlab{}.
\newblock \showarticletitle{Authentication graphs: Analyzing user behavior
  within an enterprise network}.
\newblock \bibinfo{journal}{\emph{Comput. Secur.}}  \bibinfo{volume}{48}
  (\bibinfo{year}{2015}), \bibinfo{pages}{150--166}.
\newblock


\bibitem[\protect\citeauthoryear{Kingma and Ba}{Kingma and Ba}{2014}]%
        {kingma2014adam}
\bibfield{author}{\bibinfo{person}{Diederik~P Kingma} {and}
  \bibinfo{person}{Jimmy Ba}.} \bibinfo{year}{2014}\natexlab{}.
\newblock \showarticletitle{Adam: A method for stochastic optimization}.
\newblock \bibinfo{journal}{\emph{arXiv preprint arXiv:1412.6980}}
  (\bibinfo{year}{2014}).
\newblock


\bibitem[\protect\citeauthoryear{Lee, McCormick, Neil, and Sodja}{Lee
  et~al\mbox{.}}{2019}]%
        {lee2019anomaly}
\bibfield{author}{\bibinfo{person}{Wesley Lee}, \bibinfo{person}{Tyler~H
  McCormick}, \bibinfo{person}{Joshua Neil}, {and} \bibinfo{person}{Cole
  Sodja}.} \bibinfo{year}{2019}\natexlab{}.
\newblock \showarticletitle{Anomaly detection in large scale networks with
  latent space models}.
\newblock \bibinfo{journal}{\emph{arXiv preprint arXiv:1911.05522}}
  (\bibinfo{year}{2019}).
\newblock


\bibitem[\protect\citeauthoryear{Legg, Buckley, Goldsmith, and Creese}{Legg
  et~al\mbox{.}}{2015}]%
        {legg2015automated}
\bibfield{author}{\bibinfo{person}{Philip~A Legg}, \bibinfo{person}{Oliver
  Buckley}, \bibinfo{person}{Michael Goldsmith}, {and} \bibinfo{person}{Sadie
  Creese}.} \bibinfo{year}{2015}\natexlab{}.
\newblock \showarticletitle{Automated insider threat detection system using
  user and role-based profile assessment}.
\newblock \bibinfo{journal}{\emph{IEEE Syst. J.}} \bibinfo{volume}{11},
  \bibinfo{number}{2} (\bibinfo{year}{2015}), \bibinfo{pages}{503--512}.
\newblock


\bibitem[\protect\citeauthoryear{Leichtnam, Totel, Prigent, and
  M{\'e}}{Leichtnam et~al\mbox{.}}{2020}]%
        {leichtnam2020sec2graph}
\bibfield{author}{\bibinfo{person}{Laetitia Leichtnam}, \bibinfo{person}{Eric
  Totel}, \bibinfo{person}{Nicolas Prigent}, {and} \bibinfo{person}{Ludovic
  M{\'e}}.} \bibinfo{year}{2020}\natexlab{}.
\newblock \showarticletitle{Sec2graph: Network Attack Detection Based on
  Novelty Detection on Graph Structured Data}. In
  \bibinfo{booktitle}{\emph{DIMVA}}.
\newblock


\bibitem[\protect\citeauthoryear{Lin, Chen, Cao, Tang, Zhang, Cheng, and
  Li}{Lin et~al\mbox{.}}{2018}]%
        {lin2018collaborative}
\bibfield{author}{\bibinfo{person}{Ying Lin}, \bibinfo{person}{Zhengzhang
  Chen}, \bibinfo{person}{Cheng Cao}, \bibinfo{person}{Lu-An Tang},
  \bibinfo{person}{Kai Zhang}, \bibinfo{person}{Wei Cheng}, {and}
  \bibinfo{person}{Zhichun Li}.} \bibinfo{year}{2018}\natexlab{}.
\newblock \showarticletitle{Collaborative alert ranking for anomaly detection}.
  In \bibinfo{booktitle}{\emph{CIKM}}.
\newblock


\bibitem[\protect\citeauthoryear{Liu, Wen, Zhang, Jiang, Xing, and Meng}{Liu
  et~al\mbox{.}}{2019}]%
        {liu2019log2vec}
\bibfield{author}{\bibinfo{person}{Fucheng Liu}, \bibinfo{person}{Yu Wen},
  \bibinfo{person}{Dongxue Zhang}, \bibinfo{person}{Xihe Jiang},
  \bibinfo{person}{Xinyu Xing}, {and} \bibinfo{person}{Dan Meng}.}
  \bibinfo{year}{2019}\natexlab{}.
\newblock \showarticletitle{Log2vec: A Heterogeneous Graph Embedding Based
  Approach for Detecting Cyber Threats within Enterprise}. In
  \bibinfo{booktitle}{\emph{CCS}}.
\newblock


\bibitem[\protect\citeauthoryear{Mikolov, Sutskever, Chen, Corrado, and
  Dean}{Mikolov et~al\mbox{.}}{2013}]%
        {mikolov2013distributed}
\bibfield{author}{\bibinfo{person}{Tomas Mikolov}, \bibinfo{person}{Ilya
  Sutskever}, \bibinfo{person}{Kai Chen}, \bibinfo{person}{Greg Corrado}, {and}
  \bibinfo{person}{Jeffrey Dean}.} \bibinfo{year}{2013}\natexlab{}.
\newblock \showarticletitle{Distributed representations of words and phrases
  and their compositionality}. In \bibinfo{booktitle}{\emph{NeurIPS}}.
\newblock


\bibitem[\protect\citeauthoryear{Mnih and Kavukcuoglu}{Mnih and
  Kavukcuoglu}{2013}]%
        {mnih2013learning}
\bibfield{author}{\bibinfo{person}{Andriy Mnih} {and} \bibinfo{person}{Koray
  Kavukcuoglu}.} \bibinfo{year}{2013}\natexlab{}.
\newblock \showarticletitle{Learning word embeddings efficiently with
  noise-contrastive estimation}.
\newblock \bibinfo{journal}{\emph{NeurIPS}} (\bibinfo{year}{2013}).
\newblock


\bibitem[\protect\citeauthoryear{Mnih and Teh}{Mnih and Teh}{2012}]%
        {mnih2012fast}
\bibfield{author}{\bibinfo{person}{Andriy Mnih} {and} \bibinfo{person}{Yee~Whye
  Teh}.} \bibinfo{year}{2012}\natexlab{}.
\newblock \showarticletitle{A fast and simple algorithm for training neural
  probabilistic language models}. In \bibinfo{booktitle}{\emph{ICML}}.
\newblock


\bibitem[\protect\citeauthoryear{Passino, Turcotte, and Heard}{Passino
  et~al\mbox{.}}{2020}]%
        {passino2020graph}
\bibfield{author}{\bibinfo{person}{Francesco~Sanna Passino},
  \bibinfo{person}{Melissa~JM Turcotte}, {and} \bibinfo{person}{Nicholas~A
  Heard}.} \bibinfo{year}{2020}\natexlab{}.
\newblock \showarticletitle{Graph link prediction in computer networks using
  Poisson matrix factorisation}.
\newblock \bibinfo{journal}{\emph{arXiv preprint arXiv:2001.09456}}
  (\bibinfo{year}{2020}).
\newblock


\bibitem[\protect\citeauthoryear{Paszke, Gross, Massa, Lerer, Bradbury, Chanan,
  Killeen, Lin, Gimelshein, Antiga, et~al\mbox{.}}{Paszke
  et~al\mbox{.}}{2019}]%
        {paszke2019pytorch}
\bibfield{author}{\bibinfo{person}{Adam Paszke}, \bibinfo{person}{Sam Gross},
  \bibinfo{person}{Francisco Massa}, \bibinfo{person}{Adam Lerer},
  \bibinfo{person}{James Bradbury}, \bibinfo{person}{Gregory Chanan},
  \bibinfo{person}{Trevor Killeen}, \bibinfo{person}{Zeming Lin},
  \bibinfo{person}{Natalia Gimelshein}, \bibinfo{person}{Luca Antiga},
  {et~al\mbox{.}}} \bibinfo{year}{2019}\natexlab{}.
\newblock \showarticletitle{PyTorch: An Imperative Style, High-Performance Deep
  Learning Library.}. In \bibinfo{booktitle}{\emph{NeurIPS}}.
\newblock


\bibitem[\protect\citeauthoryear{Powell}{Powell}{2020}]%
        {powell2020detecting}
\bibfield{author}{\bibinfo{person}{Brian~A Powell}.}
  \bibinfo{year}{2020}\natexlab{}.
\newblock \showarticletitle{Detecting malicious logins as graph anomalies}.
\newblock \bibinfo{journal}{\emph{J. Inf. Secur. Appl.}}  \bibinfo{volume}{54}
  (\bibinfo{year}{2020}), \bibinfo{pages}{102557}.
\newblock


\bibitem[\protect\citeauthoryear{Rashid, Agrafiotis, and Nurse}{Rashid
  et~al\mbox{.}}{2016}]%
        {rashid2016new}
\bibfield{author}{\bibinfo{person}{Tabish Rashid}, \bibinfo{person}{Ioannis
  Agrafiotis}, {and} \bibinfo{person}{Jason~RC Nurse}.}
  \bibinfo{year}{2016}\natexlab{}.
\newblock \showarticletitle{A new take on detecting insider threats: exploring
  the use of hidden markov models}. In \bibinfo{booktitle}{\emph{MIST}}.
\newblock


\bibitem[\protect\citeauthoryear{Roundy, Tamersoy, Spertus, Hart, Kats,
  Dell'Amico, and Scott}{Roundy et~al\mbox{.}}{2017}]%
        {roundy2017smoke}
\bibfield{author}{\bibinfo{person}{Kevin~A Roundy}, \bibinfo{person}{Acar
  Tamersoy}, \bibinfo{person}{Michael Spertus}, \bibinfo{person}{Michael Hart},
  \bibinfo{person}{Daniel Kats}, \bibinfo{person}{Matteo Dell'Amico}, {and}
  \bibinfo{person}{Robert Scott}.} \bibinfo{year}{2017}\natexlab{}.
\newblock \showarticletitle{Smoke detector: cross-product intrusion detection
  with weak indicators}. In \bibinfo{booktitle}{\emph{ACSAC}}.
\newblock


\bibitem[\protect\citeauthoryear{Ruder}{Ruder}{2017}]%
        {ruder2017overview}
\bibfield{author}{\bibinfo{person}{Sebastian Ruder}.}
  \bibinfo{year}{2017}\natexlab{}.
\newblock \showarticletitle{An overview of multi-task learning in deep neural
  networks}.
\newblock \bibinfo{journal}{\emph{arXiv preprint arXiv:1706.05098}}
  (\bibinfo{year}{2017}).
\newblock


\bibitem[\protect\citeauthoryear{Sexton, Storlie, and Neil}{Sexton
  et~al\mbox{.}}{2015}]%
        {sexton2015attack}
\bibfield{author}{\bibinfo{person}{Joseph Sexton}, \bibinfo{person}{Curtis
  Storlie}, {and} \bibinfo{person}{Joshua Neil}.}
  \bibinfo{year}{2015}\natexlab{}.
\newblock \showarticletitle{Attack chain detection}.
\newblock \bibinfo{journal}{\emph{Stat. Anal. Data Min.}} \bibinfo{volume}{8},
  \bibinfo{number}{5-6} (\bibinfo{year}{2015}), \bibinfo{pages}{353--363}.
\newblock


\bibitem[\protect\citeauthoryear{Shashanka, Shen, and Wang}{Shashanka
  et~al\mbox{.}}{2016}]%
        {shashanka2016user}
\bibfield{author}{\bibinfo{person}{Madhu Shashanka}, \bibinfo{person}{Min-Yi
  Shen}, {and} \bibinfo{person}{Jisheng Wang}.}
  \bibinfo{year}{2016}\natexlab{}.
\newblock \showarticletitle{User and entity behavior analytics for enterprise
  security}. In \bibinfo{booktitle}{\emph{BigData}}.
\newblock


\bibitem[\protect\citeauthoryear{Siadati and Memon}{Siadati and Memon}{2017}]%
        {siadati2017detecting}
\bibfield{author}{\bibinfo{person}{Hossein Siadati} {and}
  \bibinfo{person}{Nasir Memon}.} \bibinfo{year}{2017}\natexlab{}.
\newblock \showarticletitle{Detecting structurally anomalous logins within
  enterprise networks}. In \bibinfo{booktitle}{\emph{CCS}}.
\newblock


\bibitem[\protect\citeauthoryear{Siddiqui, Stokes, Seifert, Argyle, McCann,
  Neil, and Carroll}{Siddiqui et~al\mbox{.}}{2019}]%
        {siddiqui2019detecting}
\bibfield{author}{\bibinfo{person}{Md~Amran Siddiqui}, \bibinfo{person}{Jack~W
  Stokes}, \bibinfo{person}{Christian Seifert}, \bibinfo{person}{Evan Argyle},
  \bibinfo{person}{Robert McCann}, \bibinfo{person}{Joshua Neil}, {and}
  \bibinfo{person}{Justin Carroll}.} \bibinfo{year}{2019}\natexlab{}.
\newblock \showarticletitle{Detecting cyber attacks using anomaly detection
  with explanations and expert feedback}. In
  \bibinfo{booktitle}{\emph{ICASSP}}.
\newblock


\bibitem[\protect\citeauthoryear{Tang, Hu, and Lin}{Tang et~al\mbox{.}}{2017}]%
        {tang2017reducing}
\bibfield{author}{\bibinfo{person}{Baoming Tang}, \bibinfo{person}{Qiaona Hu},
  {and} \bibinfo{person}{Derek Lin}.} \bibinfo{year}{2017}\natexlab{}.
\newblock \showarticletitle{Reducing False Positives of User-to-Entity
  First-Access Alerts for User Behavior Analytics}. In
  \bibinfo{booktitle}{\emph{ICDM Workshops}}.
\newblock


\bibitem[\protect\citeauthoryear{Tuor, Kaplan, Hutchinson, Nichols, and
  Robinson}{Tuor et~al\mbox{.}}{2017}]%
        {tuor2017deep}
\bibfield{author}{\bibinfo{person}{Aaron Tuor}, \bibinfo{person}{Samuel
  Kaplan}, \bibinfo{person}{Brian Hutchinson}, \bibinfo{person}{Nicole
  Nichols}, {and} \bibinfo{person}{Sean Robinson}.}
  \bibinfo{year}{2017}\natexlab{}.
\newblock \showarticletitle{Deep learning for unsupervised insider threat
  detection in structured cybersecurity data streams}. In
  \bibinfo{booktitle}{\emph{AAAI Workshops}}.
\newblock


\bibitem[\protect\citeauthoryear{Tuor, Baerwolf, Knowles, Hutchinson, Nichols,
  and Jasper}{Tuor et~al\mbox{.}}{2018}]%
        {tuor2018recurrent}
\bibfield{author}{\bibinfo{person}{Aaron~Randall Tuor}, \bibinfo{person}{Ryan
  Baerwolf}, \bibinfo{person}{Nicolas Knowles}, \bibinfo{person}{Brian
  Hutchinson}, \bibinfo{person}{Nicole Nichols}, {and} \bibinfo{person}{Robert
  Jasper}.} \bibinfo{year}{2018}\natexlab{}.
\newblock \showarticletitle{Recurrent Neural Network Language Models for Open
  Vocabulary Event-Level Cyber Anomaly Detection}. In
  \bibinfo{booktitle}{\emph{AAAI Workshops}}.
\newblock


\bibitem[\protect\citeauthoryear{Turcotte, Moore, Heard, and McPhall}{Turcotte
  et~al\mbox{.}}{2016b}]%
        {turcotte2016poisson}
\bibfield{author}{\bibinfo{person}{Melissa Turcotte}, \bibinfo{person}{Juston
  Moore}, \bibinfo{person}{Nick Heard}, {and} \bibinfo{person}{Aaron McPhall}.}
  \bibinfo{year}{2016}\natexlab{b}.
\newblock \showarticletitle{Poisson factorization for peer-based anomaly
  detection}. In \bibinfo{booktitle}{\emph{ISI}}.
\newblock


\bibitem[\protect\citeauthoryear{Turcotte, Heard, and Kent}{Turcotte
  et~al\mbox{.}}{2016a}]%
        {turcotte2016modelling}
\bibfield{author}{\bibinfo{person}{Melissa~JM Turcotte},
  \bibinfo{person}{Nicholas~A Heard}, {and} \bibinfo{person}{Alexander~D
  Kent}.} \bibinfo{year}{2016}\natexlab{a}.
\newblock \showarticletitle{Modelling user behaviour in a network using
  computer event logs}.
\newblock In \bibinfo{booktitle}{\emph{Dynamic Networks and Cyber-Security}}.
  \bibinfo{publisher}{World Scientific}, \bibinfo{pages}{67--87}.
\newblock


\bibitem[\protect\citeauthoryear{Valdes and Skinner}{Valdes and
  Skinner}{2001}]%
        {valdes2001probabilistic}
\bibfield{author}{\bibinfo{person}{Alfonso Valdes} {and} \bibinfo{person}{Keith
  Skinner}.} \bibinfo{year}{2001}\natexlab{}.
\newblock \showarticletitle{Probabilistic alert correlation}. In
  \bibinfo{booktitle}{\emph{RAID}}.
\newblock


\bibitem[\protect\citeauthoryear{Veeramachaneni, Arnaldo, Korrapati, Bassias,
  and Li}{Veeramachaneni et~al\mbox{.}}{2016}]%
        {veeramachaneni2016ai}
\bibfield{author}{\bibinfo{person}{Kalyan Veeramachaneni},
  \bibinfo{person}{Ignacio Arnaldo}, \bibinfo{person}{Vamsi Korrapati},
  \bibinfo{person}{Constantinos Bassias}, {and} \bibinfo{person}{Ke Li}.}
  \bibinfo{year}{2016}\natexlab{}.
\newblock \showarticletitle{$AI^2$: training a big data machine to defend}. In
  \bibinfo{booktitle}{\emph{BigDataSecurity}}.
\newblock


\bibitem[\protect\citeauthoryear{Yen, Oprea, Onarlioglu, Leetham, Robertson,
  Juels, and Kirda}{Yen et~al\mbox{.}}{2013}]%
        {yen2013beehive}
\bibfield{author}{\bibinfo{person}{Ting-Fang Yen}, \bibinfo{person}{Alina
  Oprea}, \bibinfo{person}{Kaan Onarlioglu}, \bibinfo{person}{Todd Leetham},
  \bibinfo{person}{William Robertson}, \bibinfo{person}{Ari Juels}, {and}
  \bibinfo{person}{Engin Kirda}.} \bibinfo{year}{2013}\natexlab{}.
\newblock \showarticletitle{Beehive: Large-scale log analysis for detecting
  suspicious activity in enterprise networks}. In
  \bibinfo{booktitle}{\emph{ACSAC}}.
\newblock


\end{thebibliography}

\end{document}